\documentclass[aps, prl, twocolumn, superscriptaddress, amsmath,  tightenlines, longbibliography]{revtex4-1}

\usepackage{dcolumn}
\usepackage{graphicx}
\usepackage{mathrsfs}
\usepackage{subfigure}
\usepackage{booktabs}
\usepackage{amsmath}
\usepackage{physics}
\usepackage{dsfont}
\usepackage{amstext}
\usepackage{amssymb}
\usepackage{amsbsy}
\usepackage{bbm}
\usepackage{amsthm}
\usepackage{graphicx}
\usepackage{color}
\usepackage[colorlinks,citecolor=blue]{hyperref}

\usepackage{url}
\usepackage[colorlinks]{hyperref}
\hypersetup{%
	plainpages=true,
	breaklinks=true,       
	hypertexnames=false,  
	pageanchor=true,
	colorlinks=true,
	linkcolor={blue},
	citecolor={red},
	urlcolor={blue},
	anchorcolor={black}
}

 \makeatletter

\newcommand{\Rmnum}[1]{\expandafter\@slowromancap\romannumeral #1@}
\makeatother

\begin{document}

\title{Experimental Observation of Topological Disclination States in Lossy Electric Circuits}
\author{Jin Liu}
\thanks{These authors contributed equally}
\affiliation{School of Physics and Optoelectronics, South China University of Technology,  Guangzhou 510640, China}
\author{Wei-Wu Jin}
\thanks{These authors contributed equally}
\affiliation{School of Physics and Optoelectronics, South China University of Technology,  Guangzhou 510640, China}
\author{Zhao-Fan Cai}
\thanks{These authors contributed equally}
\affiliation{School of Physics and Optoelectronics, South China University of Technology,  Guangzhou 510640, China}
\author{Xin Wang}
\affiliation{Institute of Theoretical Physics, School of Physics, Xi'an Jiaotong University, Xi'an 710049,  China}
\author{Yu-Ran Zhang}
\affiliation{School of Physics and Optoelectronics, South China University of Technology,  Guangzhou 510640, China}
\author{Xiaomin Wei}
\affiliation{School of Physics and Optoelectronics, South China University of Technology,  Guangzhou 510640, China}
\author{Wenbo Ju}
\email[E-mail: ]{wjuphy@scut.edu.cn}
\affiliation{School of Physics and Optoelectronics, South China University of Technology,  Guangzhou 510640, China}
 \author{Zhongmin Yang}
 \email[E-mail: ]{yangzm@scut.edu.cn}
 \affiliation{School of Physics and Optoelectronics, South China University of Technology, Guangzhou 510640, China}
 \affiliation{Research Institute of Future Technology, South China Normal University, Guangzhou 510006, China}
 \affiliation{State Key Laboratory of Luminescent Materials and Devices and Institute of Optical Communication Materials, South China University of Technology, Guangzhou 510640, China}
\author{Tao Liu}
\email[E-mail: ]{liutao0716@scut.edu.cn}
\affiliation{School of Physics and Optoelectronics, South China University of Technology,  Guangzhou 510640, China}

\date{{\small \today}}


\begin{abstract}
Topological phase transitions can be remarkably induced purely by manipulating gain and loss mechanisms, offering a novel approach to engineering topological properties. Recent theoretical studies have revealed gain-loss-induced topological disclination states, along with the associated fractional charge trapped at the disclination sites. Here, we present the experimental demonstration of topological disclination states in a purely lossy electric circuit. By designing alternating lossy electric circuit networks that correspond to the disclination lattice, we observe a voltage response localized at the disclination sites and demonstrate the robustness of these states against disorder. Furthermore, we measure the charge distribution, confirming the presence of fractional charge at the disclination sites, which gives rise to the topological disclination states. Our experiment provides direct evidence of gain-loss-induced topological disclination states in electric circuits, opening new possibilities for applications in classical systems.
\end{abstract}
\maketitle

\textit{\color{blue}Introduction}.---Over the past decade, topological phases of matter have revolutionized our understanding of quantum materials and wave dynamics, introducing a transformative paradigm in modern physics \cite{Khanikaev2017,Breunig2021}. Initially rooted in condensed matter systems \cite{RevModPhys.82.3045,RevModPhys.83.1057,RevModPhys.88.035005}, these phases are characterized by robust edge states and global topological invariants. Their influence has since expanded beyond electronic systems to photonics \cite{Lu2014,RevModPhys.91.015006,Leefmans2022}, acoustic and mechanical systems \cite{Huber2016,Rocklin2017,Li2017,Ma2019}, and electric circuits \cite{Imhof2018,Lee2018,Yang2024,Zhang2025}. These advances promise technological breakthroughs in robust signal transport, disorder-resistant devices, and novel quantum computing architectures \cite{Breunig2021,Nickerson2013}.

Recent developments have further extended topology to non-Hermitian systems, which provide a natural framework for describing open dynamical systems that exchange particles and energy with their surroundings \cite{Bender2007,ElGanainy2018,Ashida2020}. The rapidly evolving field of non-Hermitian physics has uncovered a range of exotic phenomena absent in Hermitian counterparts \cite{Monifi2016,ZhangJ2018,PhysRevLett.118.045701,arXiv:1802.07964,Peng2014b,El-Ganainy2018,PhysRevLett.123.170401,PhysRevLett.123.206404,PhysRevLett.123.066405,PhysRevLett.123.206404,PhysRevLett.124.250402,Li2020,PhysRevLett.124.086801,PhysRevLett.125.186802,PhysRevB.100.054105,PhysRevB.99.235112,Zhao2019,PhysRevX.9.041015,PhysRevLett.124.056802,PhysRevB.102.235151,Bliokh2019,PhysRevB.104.165117,PhysRevLett.131.076401,PhysRevLett.127.196801,RevModPhys.93.015005,PhysRevLett.128.223903,PhysRevLett.130.103602,Zhang2022,Parto2023,Ren2022,PhysRevX.13.021007,PhysRevLett.131.116601,arXiv:2311.03777,PhysRevA.109.063329,arXiv:2403.07459,PhysRevX.14.021011,PhysRevLett.131.036402,PhysRevLett.132.050402,Leefmans2024,PhysRevLett.133.136602,arXiv:2411.10398}. These include exceptional points \cite{zdemir2019,Ding2022,Li2023}, the non-Hermitian skin effect \cite{ShunyuYao2018,PhysRevLett.125.126402,PhysRevLett.123.066404,YaoarXiv:1804.04672,PhysRevLett.121.026808,PhysRevLett.122.076801}, and unconventional bulk-boundary correspondence \cite{ShunyuYao2018,PhysRevLett.125.126402,PhysRevLett.123.066404,YaoarXiv:1804.04672}. Notably, topological phase transitions can be induced purely by manipulating gain and loss mechanisms \cite{PhysRevLett.121.213902,PhysRevLett.123.073601,PhysRevApplied.13.014047,Gao2021,PhysRevA.105.053510}, enabling a novel approach to engineering topological properties. This highlights the profound interplay between dissipation and topology, paving the way for new phases of matter beyond conventional Hermitian frameworks.

Parallel to these developments, the study of topological crystalline insulators  has demonstrated how crystalline symmetries, such as mirror and rotational symmetries, can stabilize distinct topological phases \cite{PhysRevLett.106.106802,Tang2019,Wieder2021}. Unlike conventional topological insulators, which are protected by internal symmetries like time-reversal, particle-hole, or chiral symmetry \cite{RevModPhys.88.035005}, topological crystalline insulators derive their robustness from the spatial symmetries of the crystal lattice. This leads to exotic boundary phenomena, such as hinge and corner states \cite{PhysRevB.96.245115,Benalcazar2017}, as well as unusual quantized electric responses, including fractional electric polarization and boundary-localized fractional charge \cite{PhysRevB.99.245151,PhysRevB.101.115115,Peterson2021}, which are directly linked to the crystal's geometric structure.

A particularly intriguing aspect of topological crystalline insulators  is their response to crystallographic defects, such as dislocations \cite{Ran2009,PhysRevB.97.201111,Li2018,Nayak2019,PhysRevResearch.3.033107} and disclinations \cite{PhysRevB.101.115115,PhysRevB.89.224503,PhysRevLett.124.243602,Liu2021,Geier2021,Peterson2021}. In particular, disclinations, which result from rotational misorientations in the lattice, serve as a unique platform for investigating the interplay between topology and symmetry breaking \cite{PhysRevB.101.115115,PhysRevB.89.224503}. In Hermitian cases, disclinations can trap fractional charge, providing a direct manifestation of bulk topology \cite{PhysRevB.101.115115,PhysRevB.89.224503,PhysRevLett.124.243602,Liu2021,Geier2021,Peterson2021}. Despite significant progress in both non-Hermitian topology and topological disclination states, their intersection remains largely unexplored. Non-Hermitian topological crystalline insulators establish a novel framework where crystalline symmetries and non-Hermiticity intertwine, giving rise to exotic defect-bound states governed by gain-loss mechanisms. Recent theoretical studies predict the emergence of gain-loss-induced topological disclination states \cite{PhysRevLett.133.233804}, providing a unique platform to explore non-Hermitian effects on symmetry-protected topological phases. However, experimental validation is still lacking, though promising realizations can be achieved by employing electric circuits \cite{Choi2018,Helbig2020,wu2022non,Zou2021,Hu2023,Guo2024}.

 \begin{figure*}[!tb]
 	\centering
 	\includegraphics[width=1.0\textwidth]{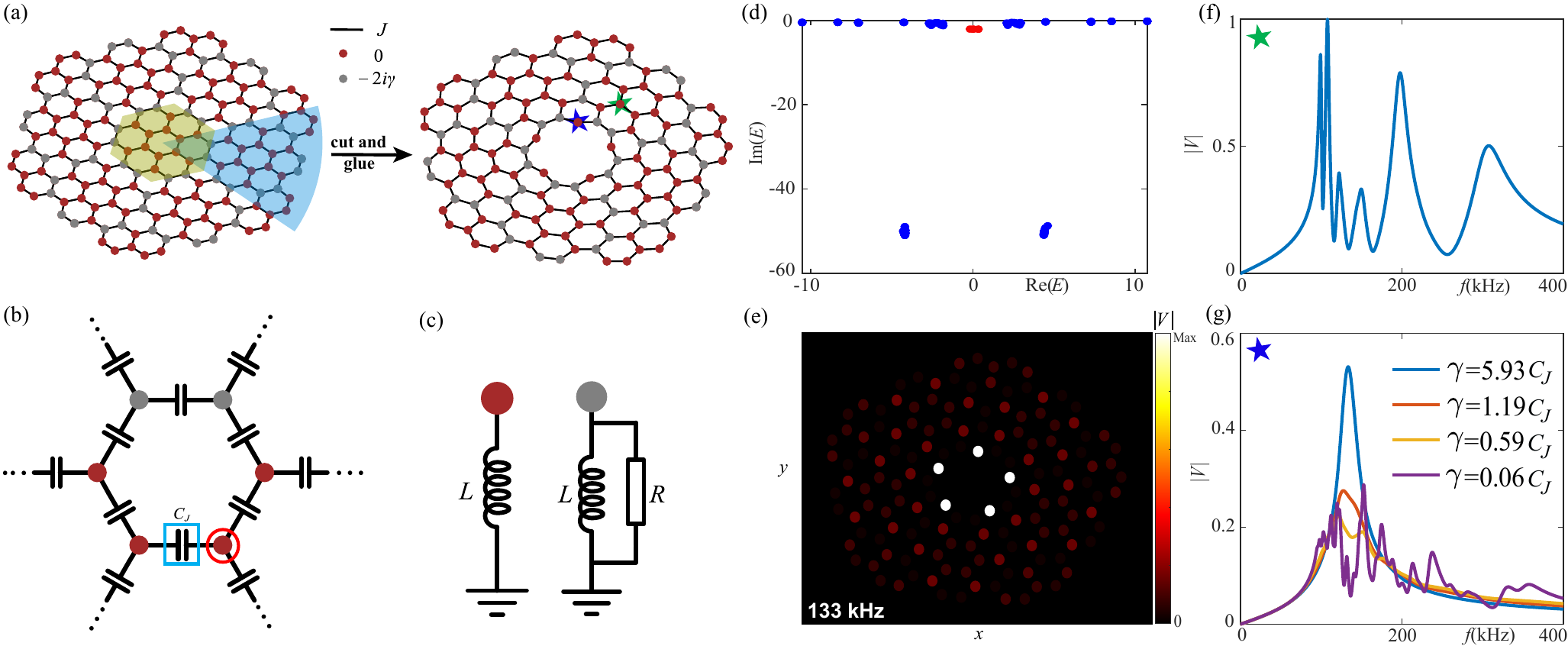}
 	\caption{(a) Schematic of a 2D non-Hermitian lattice with a disclination, created by cutting and gluing a pristine honeycomb lattice. The lattice sites represented by gray dots are dissipative, with a loss rate of  $-2\gamma$,  while the red sites are dissipationless.  The nearest-neighbor hopping (black lines) is denoted by $J$. The green and blue stars mark the bulk site and disclination site, respectively. (b) Grounded components of the electric circuit for simulating the disclination lattice in (a), where the red and gray dots represent the disclination-lattice sites within a hexagonal structure. At each red site, the corresponding node is connected to three nearest-neighbor capacitors  and one grounded inductor, as illustrated in panel (c). At each gray site, the node is connected to three nearest-neighbor capacitors,  one grounded inductor and one grounded resistor [see panel (c)], where the resistor is used to introduce onsite loss. (d) Simulated complex eigenvalue $E$, where the red dots represent five disclination states. (e) The corresponding simulated voltage distribution $\abs{V(x,y)}$ by exciting each circuit node  using  alternating voltage with the resonance frequency of $f_0=133$ kHz. Simulated voltage response $\abs{V}$ (f)  at a bulk site (green star) and  (g)  at a disclination site (blue star) for different loss $\gamma$ as a function of excitation frequency $f$. Simulation parameters are $C_J=100~\mathrm{nF}$, $R=1~\Omega$, and $L=4.7~\mathrm{\mu H}$.  }
 	\label{Fig1}
 \end{figure*}
 
In this work, we report the experimental observation of non-Hermitian topological disclination states in lossy electric circuits. Using the tight-binding model of the topological disclination lattice \cite{PhysRevLett.133.233804}, we simulate and design corresponding electric circuit networks. The gain and loss mechanisms are controlled through alternating resistors, which act as the alternating loss. By exciting each circuit node with an alternating voltage at the resonance frequency, we measure the voltage response and observe in-gap states that become localized at the disclination sites in the presence of strong loss. The robustness against disorder is verified by measuring the voltage response in a circuit where random resistors are applied at each node. Furthermore, we measure the charge distribution, confirming the fractional charge trapped at the disclination sites, which is integer for a lossless circuit. Our observations provide direct evidence of topological disclination states induced by gain-loss effects.

\textit{\color{blue}Setup}.---We begin by introducing an electric circuit model designed to simulate a non-Hermitian lattice with a disclination, as shown in Fig.~\ref{Fig1}(a). This model supports topological disclination states associated with a fractional charge of $1/2$ \cite{PhysRevLett.133.233804}.  The disclination lattice is created by cutting and reassembling a pristine honeycomb lattice [see Fig.~\ref{Fig1}(a)]. The lattice sites represented by gray dots are dissipative, with a loss rate of  $-2\gamma$,  while the red sites are dissipationless. The nearest-neighbor hopping (black lines) is denoted by $J$.

This disclination lattice can be effectively implemented using an electrical circuit network. Figures \ref{Fig1}(b,c)  show the grounded components of the electric circuit designed to simulate the disclination lattice depicted in Fig.\ref{Fig1}(a), with the red and gray dots representing the disclination-lattice sites within the hexagonal structure. At each red site, the corresponding node is connected to three nearest-neighbor capacitors with capacitance $C_J$ and one grounded inductor with inductance $L$, as shown in Fig.~\ref{Fig1}(c). At each gray site, the node is connected to three nearest-neighbor capacitors with capacitance $C_J$, one grounded inductor  with inductance $L$ and one grounded resistor with resistance $R$ (see Fig.~\ref{Fig1}(c)), where the resistor introduces onsite loss. 

Using Kirchhoff's laws \cite{Lee2018}, the disclination-lattice model in Fig.~\ref{Fig1}(a) is represented by the circuit Laplacian $J(\omega) = \mathcal{H} _{\mathrm{cir}} - E \sum_n a_n^{\dagger} a_n $ of the circuit in Fig.~\ref{Fig1}(b,c). Here, $E = 3C_J-1/(\omega^2L) $, and $\mathcal{H} _{\mathrm{cir}}$ is analogous to the Hamiltonian of the disclination model. The Laplacian relates the grounded-voltage vector  $\mathbf{V}$ to the input current vector $\mathbf{I}$ through the equation  $\mathbf{I}(\omega) = J(\omega)\mathbf{V}(\omega)$.  For the zero input current, the circuit equation  is expressed as 
\begin{align}\label{eq10}
	\mathcal{H} _{\mathrm{cir}}\boldsymbol{V}=\left( 3C_J-\frac{1}{\omega ^2L} \right) \boldsymbol{V},
\end{align}
with
\begin{align}\label{eq21}
	\mathcal{H} _{\mathrm{cir}}=-i\sum_n{\gamma _na_{n}^{\dagger}a_n}+C_J\sum_{\langle nn' \rangle } a_{n}^\dagger a_{n'}
\end{align}
where $\langle \dots \rangle$ denotes the nearest-neighbor hopping, $\gamma_n$ represents the onsite loss rate with $\gamma_n = 0$ at the red node  and  $\gamma_n=1/(\omega R)$   at the gray sites.  In this paper, otherwise specifically, we use the following experimental parameters:  $C_J=100~\mathrm{nF}$, $R=1~\Omega$, and $L=4.7~\mathrm{\mu H}$, where the resonant frequency of the circuit is calculated as $f_0=1/(2 \pi \sqrt{3LC_J}) \simeq$ 133 kHz, corresponding to $E=0$. Note that, as shown in Eqs.~(\ref{eq10}) and (\ref{eq21}), the circuit Laplacian $J(\omega)$ reflects the Hamiltonian matrix of the tight-binding lattice at the resonance frequency $\omega = \omega_0 = 2\pi f_0$.

Figure \ref{Fig1}(d) shows the simulated complex eigenvalue $E$. The in-gap states (red dots) correspond to the topological disclination states for $\gamma = 5.93 C_J$ at gray sites. The corresponding simulated voltage distribution $\abs{V(x,y)}$, obtained by exciting each circuit node with an alternating voltage at a frequency of  $f=133$ kHz, is shown in Fig.~\ref{Fig1}(e), where the in-gap states are well localized at disclination states. Note that the bulk sites and disclination sites show distinct voltage response, as shown in Fig.~\ref{Fig1}(f,g). This allows us to resolve the disclination states by applying an alternating voltage at the resonance frequency $\omega_0$.  In addition. the well-resolved voltage response at the disclination site occurs for a large loss rate $\gamma$, where the peak resonant frequency is around $f_0 = 133 $ kHz for $\gamma = 5.93 C_J$ at gray sites [see Fig.~\ref{Fig1}(g)].

\begin{figure}[!tb]
	\centering
	\includegraphics[width=8.6cm]{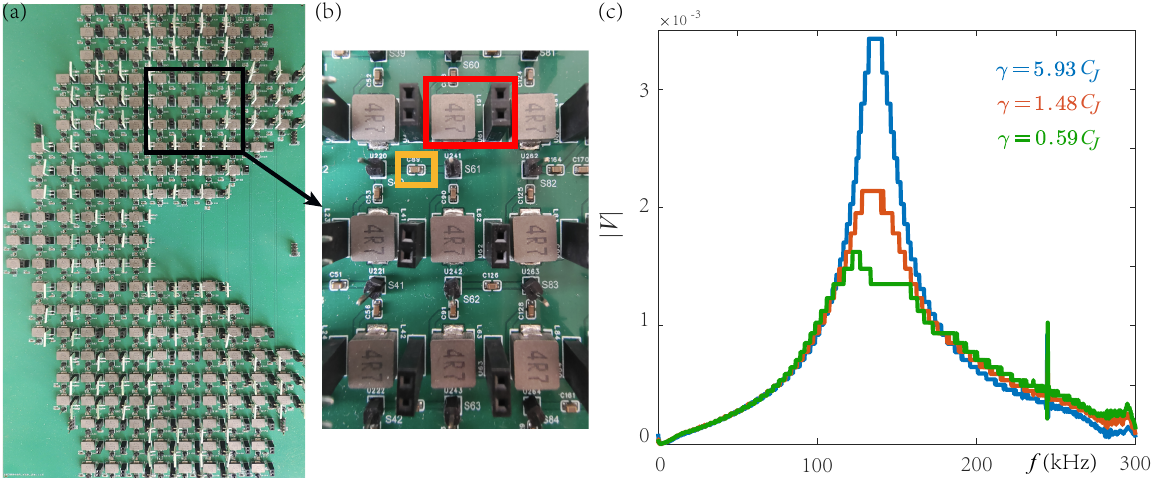}
	\caption{(a) Photograph of the printed circuit board fabricated based on the schematics in Fig.~\ref{Fig1}(b,c). (b) Enlarged view of the printed circuit board, highlighting lattice sites and nearest-neighbor reciprocal hopping with red and yellow boxes, respectively. (c) Measured voltage response $\abs{V}$ of the circuit excited at  the disclination site as a function of driving frequency $f$  for different loss rate  $\gamma$ at gray sites. The measured resonant peak is around 134 kHz.}
	\label{Fig2}
\end{figure}

\textit{\color{blue}Experimental Results}.--In order to experimentally verify gain-loss-induced topological disclination states and their associated fractional charges \cite{PhysRevLett.133.233804}, we fabricate electric circuits, with their photographs shown in Fig.~\ref{Fig2}(a,b), where the lattice site  and nearest-neighbor reciprocal hopping are highlighted by red and yellow boxes, respectively. We measure the voltage response, $\abs{V}$, of this circuit when excited at the disclination site as a function of driving frequency $f$ for different loss rates $\gamma$ at gray sites [see Fig.\ref{Fig2}(c)]. The measured resonant peak occurs around 134 kHz, which is in close agreement with the theoretical value shown in Fig.\ref{Fig1}(g). Further details about the simulation and measurement methods are provided in the Supplemental Material (SM) in Ref.~\cite{SMDisclination2025}.

\begin{figure}[!b]
	\centering
	\includegraphics[width=8.6cm]{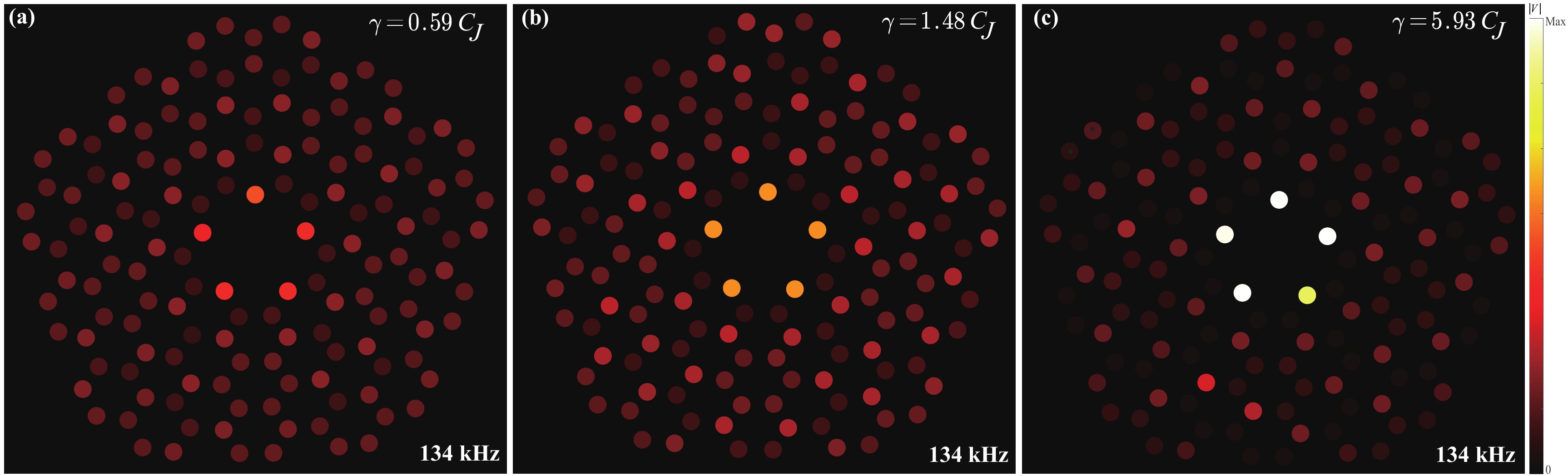}
	\caption{Experimental results of the measured voltage distribution, $\abs{V(x,y)}$, are obtained by exciting each circuit node with an alternating voltage at a frequency of $f = 134$ kHz for different loss rates $\gamma$ at gray sites: (a) $\gamma = 0.59 C_J$, (b) $\gamma = 1.48 C_J$, and (c) $\gamma = 5.93 C_J$.}
	\label{Fig3}
\end{figure}

To experimentally measure the topological disclination states, we excite each circuit node with an alternating voltage at a frequency of $f = 134$ kHz, which resonates with the disclination states. We measure the responded voltage distribution  $\abs{V(x,y)}$ for different loss rates  $\gamma$ at gray sites, as shown in Fig.~\ref{Fig3}. The disclination states cannot be clearly resolved for the weak loss rate for $\gamma = 0.59 C_J$, as shown in Fig.~\ref{Fig3}(a), which is consistent with the simulated voltage response in Fig.\ref{Fig1}(g) under weak resonance at 134 kHz. As the loss rate increases, a well-resolved disclination distribution is observed for  $\gamma = 5.93 C_J$, as shown in Fig.~\ref{Fig3}(c).

\begin{figure}[!tb]
	\centering
	\includegraphics[width=6.0cm]{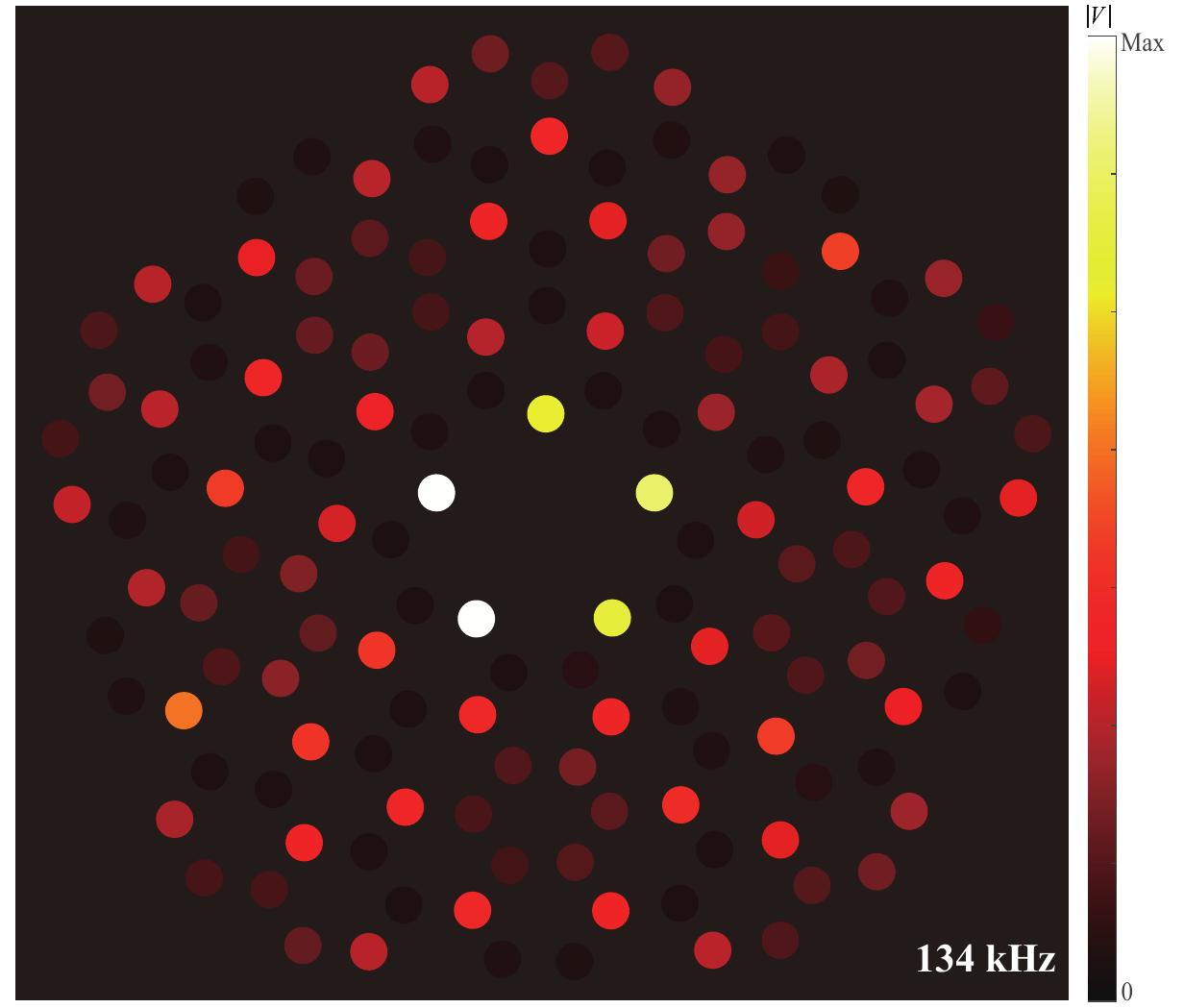}
	\caption{Experimental results of the measured voltage distribution, $\abs{V(x,y)}$, are obtained by exciting each circuit node with an alternating voltage at the frequency of $f = 134$ kHz  for $\gamma = 5.93 C_J$ at gray sites. The system is subjected to an onsite random potential by assigning a random resistor, with  its value chosen from the range $[-0.2 \gamma, ~0.2 \gamma]$, at each node. The results are averaged over 10 realizations of the disorder.}
	\label{Fig4}
\end{figure}

To assess the robustness of topological disclination states, we introduce a random onsite disorder potential by assigning a random resistor to each node. The random resistor values are uniformly drawn from the range $[-0.2 \gamma, ~0.2 \gamma]$. As shown in Fig.~\ref{Fig4}, we measure  voltage distribution $\abs{V(x,y)}$  by exciting each circuit node with an alternating voltage at a frequency of $f = 134$kHz  for $\gamma = 5.93 C_J$ at gray sites.

In Hermitian topological lattice systems, disclinations act as defects that trap fractional charges, directly reflecting the bulk topology \cite{PhysRevB.101.115115,PhysRevB.89.224503,PhysRevLett.124.243602,Liu2021,Geier2021,Peterson2021}. Interestingly, theoretical studies have shown that non-Hermitian topological disclination lattices exhibit a similar phenomenon, with fractional charge trapping serving as a hallmark of their topological nature \cite{PhysRevLett.133.233804}. 

We now move on to the experimental verification of fractional charge distribution at the disclination sites. One approach to identifying disclination charge is by analyzing the local density of states \cite{PhysRevB.99.245151,Liu2021,Peterson2021,PhysRevApplied.22.014025,PhysRevLett.133.233804}. This can be determined using the Laplacian matrix $J(\omega)$ of the circuit at the resonance frequency $\omega =\omega_0$. The Laplacian matrix $J(\omega_0)$ is reconstructed by measuring the voltage response at all lattice sites. By inverting $J(\omega_0)$, the circuit Hamiltonian  $\mathcal{H_{\text{cir}}}$ is obtained, encapsulating the key physical insights of the disclination model. Finally, diagonalizing $\mathcal{H_{\text{cir}}}$ yields the left and right eigenvectors of the circuit, with
\begin{align}\label{eq1}
	\mathcal{H} _{\mathrm{cir}}\left| \psi _{n}^{R} \right> =E\left| \psi _{n}^{R} \right> ,~\mathcal{H} _{\mathrm{cir}}^{\dagger}\left| \psi _{n}^{L} \right> =E^*\left| \psi _{n}^{L} \right>,
\end{align}
where $\left| \psi _{n}^{R} \right>$ and $\left| \psi _{n}^{L} \right>$ denote the $n$th right and left eigenvector of the circuit Hamiltonian $\mathcal{H_{\text{cir}}}$, with $\braket{\psi _{n}^{R}}{\psi _{n'}^{L}} =\delta_{nn'}$. 

The charge $Q_d$ in a specific region of the circuit can be determined as
\begin{align}\label{eq2}
	Q_d=\sum_l{\sum_n{\left| \left< \psi _{n}^{L} \middle| l \right> \left< l \middle| \psi _{n}^{R} \right> \right|~~\left( \mathrm{mod} ~1 \right)}},
\end{align}
where $l$ denotes the indices of the lattice sites, while the index $n$ represents the eigenstate corresponding to the specific eigenvalue. For bulk sites, $l$ is taken within a unit cell. For studying disclination sites, $l$ is taken within a unit cell plus the nearest neighboring disclination site. The index $n$  is taken over all occupied states.

\begin{figure}[!tb]
	\centering
	\includegraphics[width=8.6cm]{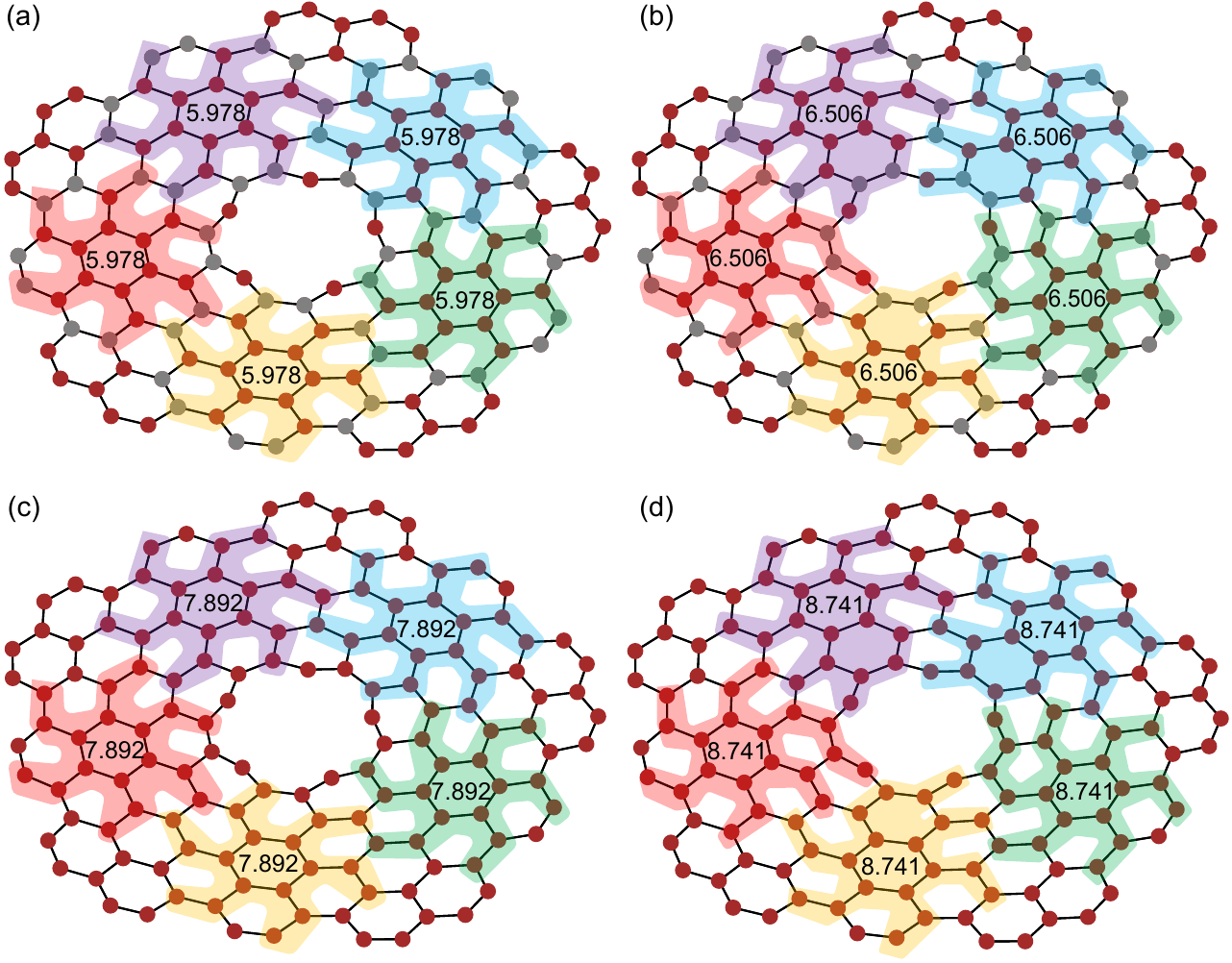}
	\caption{Charge distribution $Q_d$ (a,b) for the lossy disclination lattice with $\gamma = 5.93 C_J$ at gray sites, and (c,d) for the lossless disclination lattice   with $\gamma = 0$ at all sites.  The charge is calculated (a, c) with a unit cell, and (b, d) with a unit cell plus one disclination site.}
	\label{Fig5}
\end{figure}

Figure \ref{Fig5}(a,b) show the measured charge distribution $Q_d$  for the lossy disclination lattice  with $\gamma = 5.93 C_J$ at gray sites. In the bulk regime, the charge at each unit cell is approximately $Q_d \simeq 0$ [see Fig.~\ref{Fig5}(a)]. However, at the disclination site, the charge becomes fractional with $Q_d \simeq 0.5$ [see Fig.~\ref{Fig5}(b)]. The experimental results closely align with the theoretical ones (see details in SM \cite{SMDisclination2025}). In contrast, for the lossless circuit lattice, the charge in the bulk region at each unit cell and disclination site is approximately $Q_d \simeq 0.9$ and $Q_d \simeq 0.7$, as shown in Fig.~\ref{Fig5}(c,d).  The half-valued charge verifies the topological nature of the localized disclination states.

\textit{\color{blue}Conclusion}.---We design and simulate the electric-circuit lattice of a tight-binding model for a topological disclination lattice. The circuit is purely dissipative, with alternating resistors introduced. We excite the circuit nodes using an alternating voltage and measure the voltage response. Well-localized disclination states are observed, and their robustness against disorder is demonstrated. Furthermore, we measure the charges at both bulk and disclination sites, finding integer charges at bulk unit cells and fractional charges at disclination sites, which highlights the topological nature of the disclination states. Our electric-circuit experiments directly verify the gain-loss-induced topological disclination sites, paving the way for the exploration of novel dissipative topological phases in classical systems and potential topological applications.

\begin{acknowledgments}
T.L. acknowledges the support from  the National Natural
Science Foundation of China (Grant No. 12274142), the Fundamental Research Funds for the Central Universities (Grant No.~2023ZYGXZR020), Introduced Innovative Team Project of Guangdong Pearl River Talents Program (Grant No. 2021ZT09Z109),  and the Startup Grant of South China University of Technology (Grant No.~20210012). XW is supported by the National
Natural Science Foundation of China (NSFC) (No.~12174303). Y.R.Z. thanks the support from the National Natural Science Foundation of China (Grant No.~12475017), Natural Science Foundation of Guangdong Province (Grant No.~2024A1515010398),  and the Startup Grant of South China University of Technology (Grant No.~20240061). W.J. thanks the support from the National Natural Science Foundation of China (No.~U21A2093) and Introduced Innovative Team Project of Guangdong Pearl River Talents Program (Grant No.~2021ZT09Z109). 
F.N. is supported in part by: Nippon Telegraph and Telephone Corporation (NTT) Research, the Japan Science and Technology Agency (JST) [via the CREST Quantum Frontiers program Grant No. JPMJCR24I2, the Quantum Leap Flagship Program (Q-LEAP), and the Moonshot R$\&$D Grant Number JPMJMS2061], and the Office of Naval Research (ONR) Global (via Grant No. N62909-23-1-2074). 
\end{acknowledgments}

\textit{Note  Added}.---Upon finalizing this draft, we became aware of a recent preprint \cite{arXiv:2502.04922} that also reports topological disclination states in  gain and loss circuits.

%

\end{document}